\documentclass[aps,prl,twocolumn,showpacs,superscriptaddress,groupedaddress,floatfix]{revtex4-1}
\usepackage{graphicx}
\usepackage{amssymb}
\usepackage{amsmath}
\usepackage{verbatim}
\usepackage{dcolumn}
\usepackage{longtable}
\usepackage{multirow}
\usepackage{hyperref}
\hypersetup{ pdfborder = {0 0 0}}
\begin{document}
\title{Beyond Quantum interference and Optical pumping: invoking a Closed-loop phase}
\author{A. Kani}
\author{Harshawardhan Wanare}
\affiliation{Department of Physics, Indian Institute of Technology, Kanpur 208016, India}

\begin{abstract}
Atomic coherence effects arising from coherent light-atom interaction are conventionally known to be governed by quantum interference and optical pumping mechanisms. However, anisotropic nonlinear response driven by optical field involves another fundamental effect arising from closed-loop multiphoton transitions. This closed-loop phase dictates the tensorial structure of the nonlinear susceptibility as it governs the principal coordinate system in determining, whether the light field will either compete or cooperate with the external magnetic field stimulus. Such a treatment provides deeper understanding of all magneto-optical anisotropic response.  The magneto-optical response in all atomic systems is classified using closed-loop phase. The role of quantum interference in obtaining electromagnetically induced transparency or electromagnetically induced absorption in multi-level systems is identified.
\end{abstract}
\maketitle
A plethora of \textit{interference} based counterintuitive  phenomena have come to fore in the past two to three decades, all of which involve \textit{atomic coherence}. Electromagnetically induced transparency (EIT)~\cite{eit}, lasing without inversion (LWI)~\cite{lwi}, coherent population trapping (CPT)~\cite{cpt}, electromagnetically induced absorption (EIA)~\cite{eia}, and a family of such effects, where conventional understanding relies on a combination of basic processes involving quantum interference (QI)~\cite{quanopt} and optical pumping (OP)~\cite{OP}. These two fundamental effects are necessary and sufficient in describing the atomic response in few-level (at most three-level) atomic systems. We ask, whether mere generalization of these two effects is sufficient to describe \textit{real} multi-level atomic systems. Another pertinent aspect with regard to the role of these fundamental mechanisms arises in presence of yet another stimulus, such as an external magnetic field. A simplistic generalization of QI and OP fails to capture \textit{all} the features exhibited by such atomic systems, and we present a third central mechanism involving closed-loop multiphoton (CLMP) transitions accumulating nonlinear phase. The presence of such phase can significantly alter the expected response. Typically, one expects the \textit{optical susceptibility} to be a characteristic of a medium independent of the light field, however coherent light interaction leads to an anisotropic response~\cite{anisoatom} governed by the polarization content of the input light~\cite{kani,dual}.  We reiterate the distinction between a nonlinear response to that of an anisotropic response, the atom-light system exhibit both simultaneously. In order to determine the governing nonlinear optical susceptibility tensor $(\chi(\vec{E}))$ the induced dipole moment vector needs to be linked to the optical electric field. However, it is nearly impossible to extract a polarization dependent rank-2 tensor, unless we identify the principal coordinate system (PCS) in which $\chi(\vec{E})$ is diagonal. Without pinning down $\chi(\vec{E})$, a few measurement merely capture limited facets arising from a more complete tensorial structure. Such anisotropy plays a significant role in atom-light interaction tailored by an external magnetic field. It may be noted that the light field can either cooperate and thus reinforce the structure of the tensor $\chi(\vec{E})$ through a mere nonlinear dependence and the resulting magneto-optical effects are nonlinear generalizations at best (nonlinear magneto-optical effects~\cite{NMOE,MOEelli}), or can compete with the magnetic field stimulus to redefine the susceptibility tensor completely. The competing light field tries to align the PCS along its propagation direction, such alignment causes complex evolution as light propagates~\cite{complex} and changes conventional understanding, for example, the Voigt effect no longer results from linear-birefringence and linear-dichroism~\cite{MOE2,voigt1,voigt2}. The closed-loop phase governs this competition or cooperation between the two in determining the optical susceptibility, and further allows us to classify \textit{all} atomic systems in their anisotropic magneto-optical response. We address specifically atomic systems interacting with a monochromatic optical field in the presence of an arbitrarily oriented magnetic field. We first discuss independently light induced anisotropy (LIA), followed by magnetic field induced anisotropy (MIA) and then understand the anisotropy involving them simultaneously.

\begin{figure}
\includegraphics[width=\linewidth]{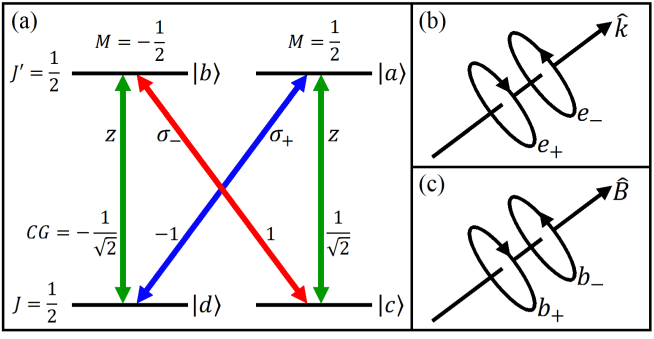}
\caption{\label{J1}(a) The four-level atomic system in the atomic spherical basis $\{\sigma_+,\sigma_-,\hat{z}\}$. The electric field components $(E_+,E_-,E_z)$ couple the Zeeman sub-levels in accordance to the dipole selection rule. (b) and (c) are the principal axes dictated by the light and the magnetic field, respectively.}
\end{figure}

{\bf{LIA: }}The symmetry in the Clebsch-Gordan $(CG)$ coefficients governing  the dipole transitions ensures that the atomic media are intrinsically optically isotropic. However, an elliptically polarized light interacting with atom can break the symmetry of the atomic response and induce anisotropy~\cite{PSR1,PSR2}. As most atomic systems exhibit anisotropy when excited with elliptically polarized light~\cite{PSR2}, we consider an arche-typical system which displays LIA. The atomic system consists of four levels with the ground and excited states comprising of states with the total angular momentum $J=J'=1/2$, interacting with a monochromatic light field as shown in Fig.~\ref{J1}(a). In order to describe the light matter interaction, defining the quantization axis $(\hat z)$ is imperative. When the energy levels  in each manifold are degenerate, one is free to choose the quantization axis without being constrained in any manner by the atomic system. We utilize this freedom to identify the PCS in which the nonlinear susceptibility tensor becomes apparent. A choice, along the direction of propagation of light, namely $\hat k$, turns out to be the PCS. Under this choice the analytical solution of the density matrix elements that correspond to the induced dipole moment after undertaking the rotating wave approximation are,
\begin{eqnarray}\label{indud}
&&\rho_{ad}=\alpha|\Omega_{bc}|^2\Omega_{ad}e^{-i\omega t},\qquad \rho_{ac}=0,\nonumber\\
&&\rho_{bc}=\alpha|\Omega_{ad}|^2\Omega_{bc}e^{-i\omega t},\qquad\rho_{bd}=0,\\
&&\alpha=\frac{2(2 \delta -i \gamma )}{\left(\gamma ^2+4 \delta ^2\right)(|\Omega_{ad}|^2+|\Omega_{bc}|^2)+16|\Omega_{ad}|^2|\Omega_{bc}|^2}\nonumber,
\end{eqnarray}
where, $\omega$ is the frequency of light, $\Omega_{ij}$ are the Rabi frequencies, $\gamma$ is the spontaneous emission rate, and $\delta$ is the detuning. The above relations clearly indicate that the induced dipole moment components are proportional to the corresponding applied electric field components through their Rabi frequencies, and the associated basis $\{\sigma_+,\sigma_-,\hat z\text{ along }\hat{k}\}$ forms the PCS for expressing the nonlinear susceptibility tensor. Eqn.~\eqref{indud} is obtained from the Master equation written in the principal basis, thus, the PCS is identified with the light field ($\{e_+,e_-,\hat k\}$) and is shown in Fig.~\ref{J1}(b). The above result is not perturbative~\cite{multipole1,MOEelli}, and captures the \textit{nonlinear} response. Since the light does not induce dipole moment along the $\hat k$, the two-dimensional susceptibility matrix would suffice.

The LIA arises due to the imbalance in the eigen polarizations $\{e_+,e_-\}$ (left and right circular polarizations) leading to OP. These circularly polarized components experience different absorption (refractive indices) resulting in circular-dichroism (circular-birefringence). The polarization rotation due to LIA is also known as Polarization Self Rotation~\cite{PSR1,PSR2}. In contrast, under a different choice of quantization axis the anisotropy can be understood as arising from QI~\cite{kani}. In general, both these mechanisms can contribute to LIA. However, when the energy levels are degenerate the OP and the QI mechanisms are indistinguishable in the steady state, since QI populates the coherent superposition states, moreover, QI mechanism is intrinsically subnatural.

{\bf MIA: }In presence of an external magnetic field all systems become optically anisotropic. In order to understand purely the role of magnetic field, in contrast to light, we consider a system that precludes any OP mechanism discussed above. Atomic  systems  which do not have any ground state degeneracy are  most amenable to such a treatment, for example  a four level system with the total angular momentum of the ground and excited states being $J=0$ and $J\rq{}=1$~\cite{MIA1}. However, the excited state QI can itself lead to LIA. To exclude such LIA one has to work with a weak electromagnetic field $(\Omega_{ij}\ll\gamma)$, wherein the excited state zeeman coherence is not effective~\cite{weak}, and study the effect as solely arising from magnetic field.

Choosing the quantization axis along the direction of the magnetic field $(\hat B)$ is natural. In the weak excitation regime the medium is linear, and one can extract the linear susceptibility tensor, and thus identify the PCS. The  principal axes associated with MIA are $\{b_+,b_-,\hat B \}$ ($\{\sigma_+,\sigma_-,\hat z\text{ along }\vec{B}\}$) as shown in Fig.~\ref{J1}(c). The magnetic field induces anisotropy by lifting the degeneracy of the energy eigenvalues. The magneto-optical effects depend on the relative orientation of $\hat B$ and $\hat k$, i.e. whether the light propagates along (Faraday effect) or transverse (Voigt effect) to the direction of the magnetic field~\cite{MOE2}. The light propagation associated with Faraday (Voigt) effect is circular-birefringence and circular-dichroism (linear-birefringence and linear-dichroism)~\cite{faraday1,faraday2,voigt1,voigt2}. 

\begin{figure*}[ht]
\includegraphics[width=\textwidth]{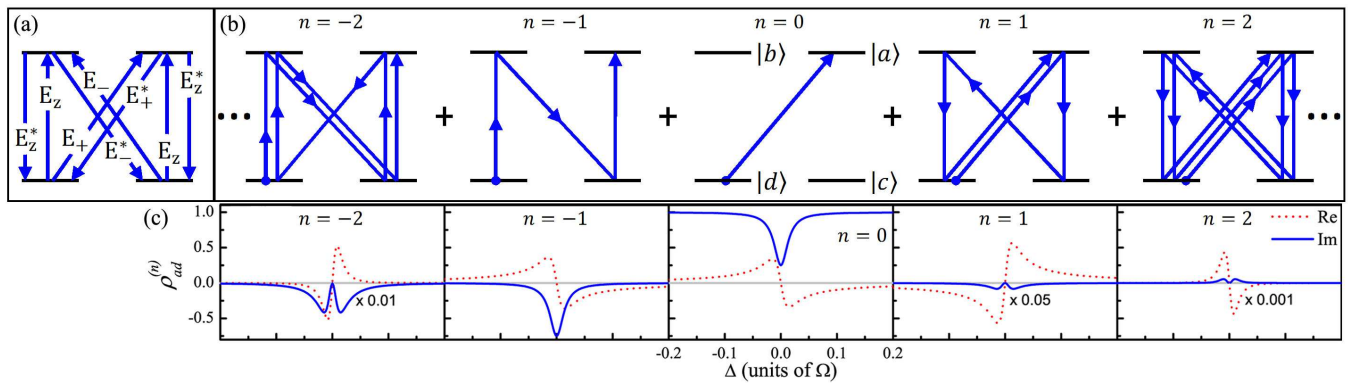}
\caption{\label{multi}(a) The single-photon transitions governed by the phase of the corresponding coupling field. (b) All possible multiphoton excitation pathways for $|d\rangle\rightarrow|a\rangle$ transition. (c) The coefficients $\rho_{ad}^{(n)}$ as a function of magnetic field ($\Delta=\mu_B B/\hbar$) with $\Omega=\sqrt{\sum\Omega_{ij}^2}=\gamma/100$, for $\epsilon=30^{\circ}$ under the Voigt configuration, where n indicates the corresponding multiphoton loop given in Eqn.~\eqref{phase} and ellipticity of light polarization is defined as $\tan(\epsilon)$.}
\end{figure*}

{\bf LIA and MIA: }Atomic systems that exhibit anisotropy arising from both the light field and the magnetic field simultaneously, involve competition in defining the PCS. The external magnetic field tends to align the quantization axis $\hat{z}$ along $\hat{B}$~\cite{Stern_Gerlach}. In contrast, is there a compelling reason for the PCS being set by the light field? Indeed, we identify a phase accumulated along the underlying CLMP pathways as the central piece of the puzzle.

In order to bring forth the importance of CLMP transitions, we separate the applied field phases explicitly as arising in the diagonal and off-diagonal elements of the density matrix as
\begin{equation}\label{phase}
\begin{aligned}
\rho_{ii}=&\sum _{n=-\infty}^{+\infty}\rho_{ii}^{(n)}\eta^n,&\text{(population)}\\
\rho_{ij}=&\sum _{n=-\infty}^{+\infty}\rho_{ij}^{(n)}e^{i(\phi_+-\phi_z)}\eta^n,&\text{(Zeeman coherence)}\\
\rho_{ij}=&\sum _{n=-\infty}^{+\infty}\rho_{ij}^{(n)}e^{i(\phi_{ij}-\omega t)}\eta^n,&\text{(dipole moment)}
\end{aligned}
\end{equation}
where, $\phi_{ij}$ is the phase $(\phi_+,\phi_-,\phi_z)$ of the electric field component $(E_+,E_-,E_z)$ coupling the levels $|j\rangle \to |i\rangle$, and the close-loop phase $\eta =e^{i(\phi_++\phi_--2\phi_z)}$. This closed-loop phase $(\eta)$ governs the induced dipole moment, as apparent in the above decomposition, for which the $\hat{z}$ axis is taken along the $\hat{B}$ field direction. As the dipole moment is governed by light polarization dependent nonlinear phase $(e^{i\phi_{ij}}\eta^n)$, the magnetic field alone is not sufficient to establish the PCS that will diagonalize the susceptibility. However, in the absence of magnetic field, it is this nonlinear phase or the corresponding light field that defines the PCS. This nonlinear phase arises because the light drives the atom along CLMP pathways, for example, the phase accumulation along excitation pathways $|d\rangle\rightarrow|a\rangle$ are shown pictorially in Fig.~\ref{multi}(b), along the lines of Eqn.~\eqref{phase}. The magnetic field determines the weight factors $(\rho_{ij}^{(n)})$ that governs the interference of the various pathways (Fig.~\ref{multi}(c)). In essence, the light drives the CLMP pathways trying to establish the PCS along $\{e_+,e_-,\hat k\}$, and simultaneously the magnetic field tries to establish the PCS along $\{b_+,b_-,\hat B\}$ by curtailing the effect of the light field. This allows us to classify the interplay into two categories, one involving a \textit{competition} between the light and magnetic field in determining the PCS and the other involving the light field \textit{reinforcing} the magnetic field PCS. Essentially, it is governed by the Eqn.~\eqref{phase}, and further constrained by $\eta$ and the finiteness of the coefficients $\rho_{ij}^{(n\neq0)}$. Table~\ref{tab} contains the summary of the possibilities.  We now present a few special cases of atomic systems, polarization content of the light field, and $\vec{B}-\vec{k}$ direction dependence in detail.

\begin{table*}[ht]
\caption{\label{tab}Classification of \textit{all} atomic transitions driven by coherent light field in the presence of magnetic field. }
\begin{ruledtabular}
\begin{tabular}{ccccccc}
Category & Closed-loop & Phase& Coefficients & Interplay & Specific transitions & Specific conditions for $\forall$ $J\rightarrow J'$ transitions\\
\hline
1 & Absent & - & - & Cooperation & $J=0\leftrightarrow J=1$ & Faraday configuration\\
2 & Present & $\eta=1$ & - & Cooperation & $J=1\rightarrow J=1$ & Coupled to linearly polarized light \\
3 & Present & $\eta\neq1$ & $\rho_{ij}^{(n\neq0)}=0$ & Cooperation &  &  For large B field, beyond the subnatural strength  \\
4& Present & $\eta\neq1$ & $\rho_{ij}^{(n\neq0)}\neq0$ & Competition & & All transitions other than listed above\\
\end{tabular}
\end{ruledtabular}
\end{table*}

\begin{figure*}[ht]
\includegraphics[width=\textwidth]{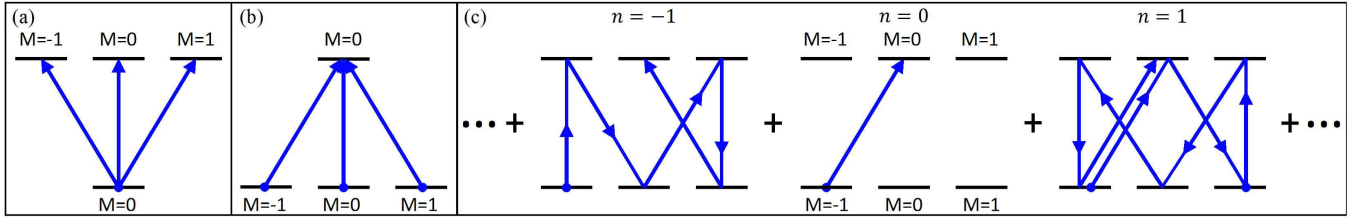}
\caption{\label{iso} (a) and (b) Atomic systems $J=0\to J'=1$ and $J=1\to J'=0$ show open (delinked) pathways. (c) For atomic system $J=1\to J'=1$, the multiphoton excitation pathways contributing to the $|M=-1\rangle\to|M=0\rangle$ transition are shown. Note that all the excitation pathways result in an overall linear phase dictated by only the $E_+$ component.}
\end{figure*}
\begin{figure}[ht]
\includegraphics[width=\linewidth]{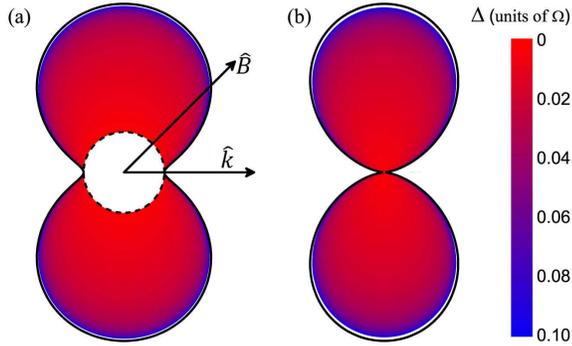}
\caption{\label{abs}Radial plot of the resonant absorption as a function of $\hat{B}$ field direction (a) for elliptically polarized light $(\epsilon=30^\circ)$ with the major axis in the $\hat B-\hat k$ plane and (b) for circularly polarized light $(\epsilon=45^\circ)$. The colour map indicates the strength of the $B$-field and the extreme cases are marked with dashed line $(\Delta=0)$ and solid line $(\Delta=\Omega)$.}
\end{figure}

Atomic transitions involving $J=0\rightarrow J'=1$ and $J=1\rightarrow J'=0$ involve delinked single-photon transitions as shown in Fig.~\ref{iso}(a and b), and do not contain multiphoton pathways that can accumulate nonlinear phase (category-1). Atomic system with $J=1\rightarrow J'=1$, involve CLMP pathways, yet contribute a trivial closed-loop phase $\eta=1$, and arises specifically from the selection rule $M_J=0\not\leftrightarrow M_{J'}=0$ not being allowed (category-2, Fig.~\ref{iso}(c)). In these systems, light field provides no competition or constraint via $\eta$. These media continue to remain isotropic for zero magnetic field, irrespective of the polarization state of the light field. In the presence of magnetic field, OP and QI can lead to LIA, which invariably reinforces MIA. Even though a few specific systems exhibiting isotropy at zero magnetic field are understood~\cite{jdeplia,PSR2}, the isotropy arises fundamentally due to the absence of the closed-loop phase.

The basic configuration that leads to competition between the light and magnetic field involves the closed-loop four-level unit shown in Fig.~\ref{multi}(a). This element is common to \textit{all} the atomic systems that exhibit LIA. However, when the atomic systems are coupled to linearly polarized light, $\eta=1$ regardless of the direction of the polarization and thus fall in category-2. Furthermore, in the Faraday configuration for any polarization state, the closed-loop pathways get de-linked since $E_z=0$ (category-1). In these cases the resulting MIA is reinforced by the nonlinear light interaction.

Other than the above \textit{few cases} the light and magnetic field compete to determine the PCS.  It is here that the competition brings added complexity as the PCS depends on the polarization state as well as strengths of the two fields, and their relative orientation. However, irrespective of the above, a strong magnetic field can make the CLMP transitions insignificant $(\rho_{ij}^{(n\neq0)}\to0)$ as seen in Fig.~\ref{multi}(c), and overwhelm the light in defining the PCS (category-3).  Since the CLMP transitions are subnatural the competition between the two occurs typically for magnetic field strengths within the subnatural line width associated with QI.  And the competition transforms the PCS governing the nonlinear susceptibility tensor from $\{e_+,e_−,\hat{k}\}$ to $\{b_+,b_−,\hat{B}\}$ as the magnetic field strength is increased from zero beyond the subnatural field strength. Such alignment of PCS by the magnetic field plays a central role under the Voigt configuration, since the physics of light propagation changes from circular-birefringence and circular-dichroism (at $\vec{B}=0$) to linear-birefringence and linear-dichroism (at large $\vec{B}$). Within these two extremes the complex evolution of light polarization as it propagates has been calculated by solving the coupled Maxwell-Bloch equation~\cite{masmax1,masmax2}, and is shown in Fig.~\ref{abs}.

\begin{figure}[!t]
\includegraphics[width=\linewidth]{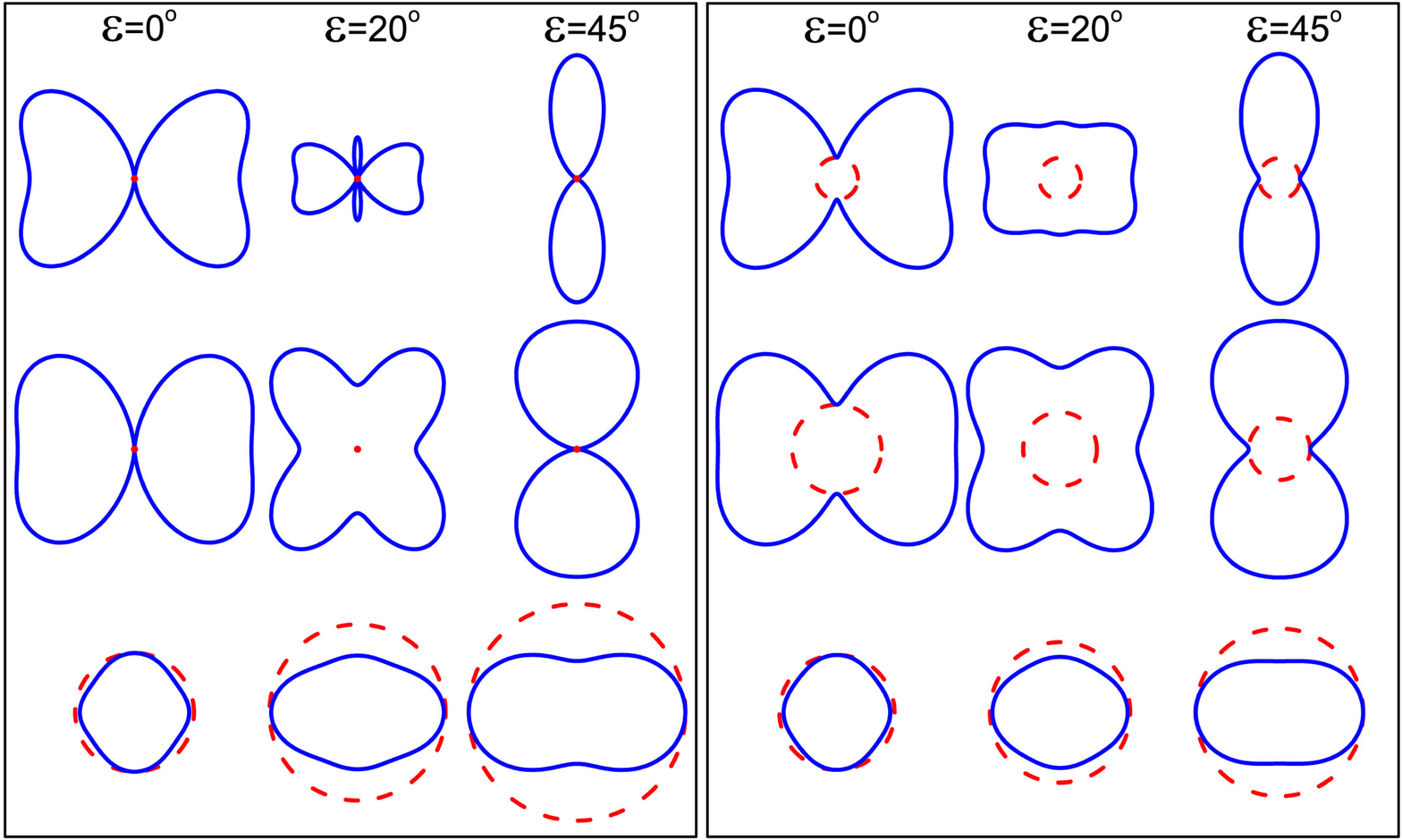}
\caption{\label{EITEIA1}The direction dependent radial plot of the absorption (similar to Fig.~\ref{abs}) for two extreme cases (dashed line ($\Delta=0$) and solid line ($\Delta=\Omega$)) for the atomic systems $J=2\to J'=1,2,3$ (top row to bottom row) coupled to different polarization whose major axis lies in the $\hat{B}-\hat{k}$ plane, without (left) and with (right) ground state decoherence $(\gamma_d=\Omega/1000)$.}
\end{figure}

Even though atomic systems involve both OP and QI mechanisms, understanding the effect of QI becomes crucial, because depending on the atomic level structure the QI plays a central role in enhancing or reducing the absorption via EIA or EIT. There are numerous experiments and theoretical analyses on several atomic systems which tend to classify them in terms of the resulting QI effects~\cite{eiteia1,eiteia2,eiteia3}. However, in order to understand the role of QI one would need to differentiate the effects of OP and QI at zero magnetic field. The two extreme cases, one (at large $\vec{B}$) involving dominantly OP and other ($\vec{B}=0$) involving both OP and QI allows us to differentiate their roles. As we have obtained the PCS in these two extremes, we identify and pin down the role of QI in EIA or EIT. At sufficiently large magnetic field the QI features are lost and the OP mechanism drives the atomic system. The OP mechanism itself strongly depends on the polarization state of the light field and the direction of light propagation $\hat{k}$ with respect to $\hat{B}$ field, and the associated atomic absorption is shown in Fig.~\ref{EITEIA1}. As the magnetic field strength is reduced, the QI effects comes into play to make the physics invariant across various directions of light propagation, by either enhancing or reducing the OP response. As shown in Fig.~\ref{EITEIA1}, for the system $J=2\to J'=3$, there is an enhanced absorption, where as $J=2\to J'=1,2$ there is a reduced absorption at zero B. The extent of change in the absorption depends on the $\hat{B}-\hat{k}$ orientation and polarization~\cite{eiteia2}. It is this direction dependent global picture (Fig.~\ref{EITEIA1}) that allows us to identify the role of QI resulting in EIT or EIA. We further note that for finite B-field the role of QI is negligible and OP drives the direction dependent absorption. Even in absence of QI such direction dependent OP response has been associated with EIT or EIA~\cite{dirl1,dirl2,dirl3}.

We have also studied the effect of ground state decoherence, which can not be ignored at finite temperature. The decoherence destroys both the OP and QI effects, more importantly it destroys the contribution of CLMP transitions by depleting the magnitude of $\rho_{ij}^{(n\neq 0)}$, and equalizes the single-photon $\rho_{ij}^{(0)}$ for all the polarization components. The effect of decoherence on the optical response is shown in Fig.~\ref{EITEIA1}(right panel). For decoherence rates $(\gamma_d)$ larger than the Rabi frequency the CLMP contributions become negligible.

In conclusion, we identify the phase accumulation along the CLMP transitions as the basic mechanism behind fixing the PCS, and explain the interplay of light field in altering the magneto-optical response, which allows us to reformulate the conventional understanding of magneto-optical effects. With the help of the nonlinear susceptibility tensor, we understand the role of QI in multi-level systems resulting in EIT or EIA. We present a classification of magneto-optical effects in Table I, and show the importance of these considerations for small magnetic field within the subnatural line widths. Vectorial magnetometry with high sensitivity and dynamic range could be realized exploiting the closed-loop phase.

\end{document}